\pdfoutput=1
\documentclass[sigplan,10pt]{acmart}
\renewcommand\footnotetextcopyrightpermission[1]{}
\usepackage{xspace} 
\usepackage{enumitem} 
\usepackage{multirow} 
\usepackage{graphicx} 
\usepackage{subcaption} 
\usepackage{listings} 
\usepackage{hyperref} 
\usepackage{xcolor} 
\usepackage{caption}
\usepackage{algorithm}
\usepackage{algpseudocode}

\usepackage{import}
\usepackage{xifthen}
\usepackage{pdfpages}
\usepackage{transparent}
\usepackage{amsmath}

\newcommand{\sys}{\textsc{HMTRace}\xspace}
\newcommand{\tsan}{\textit{Thread sanitizer}\xspace}
\newcommand{\archer}{\textit{Archer}\xspace}
\newcommand{\mte}{\textit{MTE}\xspace}

\algnewcommand{\LeftComment}[1]{\Statex \(\triangleright\) #1}
\algnewcommand{\LeftCommentE}[1]{\Statex  #1}

\title{\sys: Hardware-Assisted Memory-Tagging based Dynamic Data Race Detection}

\author{Jaidev Shastri}
\email{jaidevshastri@vt.edu}
\affiliation{%
	\institution{Virginia Tech}
	\country{USA}
}

\author{Xiaoguang Wang}
\email{xgwang9@uic.edu}
\affiliation{%
	\institution{University of Illinois Chicago}
	\country{USA}
}

\author{Basavesh Ammanaghatta Shivakumar}
\email{basavesh@vt.edu}
\affiliation{%
	\institution{Virginia Tech}
	\country{USA}
}

\author{Freek Verbeek}
\email{freek@vt.edu}
\affiliation{%
	\institution{Virginia Tech \& Open University}
	\country{USA \& The Netherlands}
}

\author{Binoy Ravindran}
\email{binoy@vt.edu}
\affiliation{%
	\institution{Virginia Tech}
	\country{USA}
}

\begin{document}

\begin{abstract}
Data race, a category of insidious software concurrency bugs, is often challenging and resource-intensive to detect and debug. Existing dynamic race detection tools incur significant execution time and memory overhead while exhibiting high false positives. This paper proposes \sys, a novel Armv8.5-A memory tag extension (\mte) based dynamic data race detection framework, emphasizing low compute and memory requirements while maintaining high accuracy and precision. \sys supports race detection in userspace \textit{OpenMP} and \textit{Pthread} based multi-threaded \textit{C} applications. \sys showcases a combined f1-score of \textbf{0.86} while incurring a mean execution time overhead of \textbf{4.01\%} and peak memory (RSS) overhead of \textbf{54.31\%}. \sys also \textbf{does not report false positives}, asserting all reported races.
\end{abstract}

\maketitle
\pagestyle{plain} 

\section{Introduction}


The past decade has seen a tremendous shift in the programming paradigm, from sequential single-threaded to data- and task-parallel execution models, to the extent that legacy and newer programs need to leverage processor's multi-threading capability in both HPC and client space alike~\cite{ParallelismSurvey}. Amongst the parallel execution models, the shared memory execution model is the de facto for data and task parallelism. Shared memory programming models such as OpenMP~\cite{OpenMP} and Pthread~\cite{Pthread} are defined in comprehensive specifications, providing the developers with an exhaustive set of directives and APIs to facilitate the development of multi-threaded programs. However, due to the complexities of such specifications and API usage, lack or misuse of synchronization directives, unintended sharing of memory allocations, architectural memory ordering constraints, or lack of codebase expertise, data race bugs are inadvertently introduced, thereby deviating from a program's functional correctness.

Data race~\cite{DataRace} is a class of software concurrency bugs that is often challenging and compute-intensive to detect and debug due to the inherent non-determinism in parallel programming models. This non-determinism is primarily due to numerous ways of interleaving thread executions within a multi-threaded program. Precise data race detection is a NP-hard problem~\cite{DataRaceComplexity}, requiring compute nodes with sufficiently high processor count and memory bandwidth to detect and debug data races effectively. Often, the execution of the program under the influence of a race detection framework can lead to changes in its execution state due to the overhead introduced by the framework. For example, a latency-sensitive web server may incur timeouts due to delays in socket operations. \textit{Data race} can be defined as a program's incorrect execution state resulting from concurrent access to a shared memory location by multiple threads, with at least one of the threads performing a write operation on the shared memory. 

Data race detection frameworks can be classified as static, which performs data race analysis offline~\cite{RacerD, LLOV, O2, OpenRace, OMPRacer, Cord, Racemob}; dynamic, which detects data race during program execution~\cite{Fastrack, Pacer, Tsan, Archer, Romp, DynamicO1}; or hybrid, which performs analysis during runtime and offline~\cite{Racemob, Racez, Prorace}. Static race detection frameworks primarily suffer from the state space explosion problem~\cite{StaticDataRace} and low specificity due to reported false positives, thereby limiting their scalability. While typically less accurate than most static frameworks, dynamic race detection frameworks provide a practical solution for data race detection by reporting significantly lower false positives with reasonable performance overhead. This practicality is evident from the popularity of mainstream frameworks such as \tsan~\cite{Tsan} and \archer~\cite{Archer}, which are a part of the mainline LLVM~\cite{LLVM:CGO04} compiler codebase. 

Reporting lower false positives improves the reliability of a race detection framework for assessing and triaging reported races, with \textit{zero false positives} (specificity = 1) asserting the presence of a data race. Lower specificity correlates to an increase in a developer's race assessment and triaging effort, which can be significant in large-scale projects. While the current mainstream dynamic race detection frameworks have significantly lowered their false positives~\cite{DRB} through improvements in their race detection algorithms or approaches, they could still report a few false positives (even when not sacrificing the accuracy of reported data races for performance) — this potential for false positives necessitates additional effort from the developer to validate the reported race before triaging. 

Most state-of-the-art dynamic race detection frameworks leverage happens-before~\cite{HBR} and lockset~\cite{LocksetAnalysis} analyses to infer data races. However, logical vector-clock-based happens-before analysis for data race detection incurs significant execution time overhead (processor and memory) and may result in false negatives due to metadata overwrites. The overheads introduced by instrumentation in non-hardware-based approaches may make dynamic race detection less feasible in latency-sensitive and release environments. Furthermore, dynamic race detection frameworks are typically run alongside fuzzing frameworks during development phases, where an increase in the execution time of individual code segments (due to instrumentation) directly translates to a lower code coverage rate, thereby impacting the overall effectiveness of the fuzzing framework. In instances where programs are deployed in cloud environments through IaaS~\cite{IaaS} for validation and release phases, complexity in race detection may result in resource overheads, which in turn translate to an increase in the operation cost of the cloud infrastructure~\cite{IaasCost}.

This paper presents \sys: hardware-assisted memory-tagging based race detection. This dynamic race detection framework leverages Arm-v8.5-A memory tagging extension (\mte)~\cite{MTEWhitepaper} to detect data races with accuracy and precision comparable to current state-of-the-art dynamic race detection frameworks while incurring significantly lower performance overhead. \mte provides architectural capabilities for the processor to trace access to allocations from pointers through memory tagging~\cite{TaggedArch}. The design of \sys stems from an observation that all shared memory accesses in shared programming models, such as \textit{OpenMP} and \textit{Pthread}, can be performed through pointers. This vital observation provides a backdrop for leveraging \mte to restrict access to shared allocations and detect potential data races. \sys instruments target binary to infer a \textit{happens-before-like} relationship between memory events to assert data races. We term this happens-before-like inference as \textit{Tag-based race inference} (TBRI). While \textit{TBRI} is not strong enough to infer causality between memory events (more than two), it is sufficient to detect data race as the framework propagates necessary information between related memory events to detect concurrent memory access.

This paper's primary contribution is a hardware-assisted approach to dynamically detecting data races in the global, heap, and stack segments of \textit{OpenMP-} and \textit{Pthread-}based C multi-threaded applications. 

We make the following contributions:
\begin{itemize}[topsep=0pt,after=\vspace{\baselineskip}]
	\item We design \sys, a hardware-assisted dynamic race detection framework leveraging lightweight \textit{Arm-v8.5-A memory-tagging} (\mte) and \textit{lockset analysis} to assert data races. We formally describe the \textit{tag-based race inference} algorithm to infer conditions similar to happens-before for race detection using \mte. 
	\item We implement \sys as a series of analysis and instrumentation passes on llvm-16.0.6~\cite{LLVM:CGO04}, with a context-sensitive inter-procedural pointer analysis implemented by extending SVF~\cite{SVF}, to identify the scope of instrumentation for data race detection accurately. We also implement kernel patches to set protection attributes for the global, heap, and stack segments to facilitate memory tagging. 
	
	We aim to open-source our codebase with evaluation datasets upon publication to foster future research, transparency, and reproducibility.
	\item We evaluate the statistical performance (accuracy, precision, and f1-score) of \sys against \tsan~\cite{Tsan} and \archer~\cite{Archer} using DataRaceBench (DRB) (v1.4.1)~\cite{DRB}. We compare the execution time overhead of \sys against \tsan and \archer on \textit{Pixel 8's Tensor G3} cores using benchmarks such as DRB, Redis~\cite{Redis}, Nginx~\cite{Nginx}, Sqlite~\cite{Sqlite}, Memcached~\cite{Memcached}, Pigz~\cite{Pigz}, Aget~\cite{Aget}, Apache Httpd~\cite{ApacheHttpd}, Parsec/Splash-2x~\cite{Parsec,ParsecRepor} (C based).
\end{itemize}

 \sys detects data races with an accuracy of \textit{0.88} and a precision of \textit{1.0} when executing the DataRaceBench v1.4.1 suite. \sys incurs a mean execution time overhead of \textit{4.01\%} and a peak memory (RSS) overhead of \textit{54.31\%} when executing large-scale applications. In comparison, state-of-the-art dynamic race detection framework, such as \archer, exhibits an accuracy of \textit{0.86} and a precision of \textit{0.97} while incurring a mean execution time overhead of \textit{356.94\%} and a peak memory overhead of \textit{500.19\%}. \sys does not report false positives, asserting all reported data races.

\section{Background}
In this section, we provide the necessary background on the happens-before analysis~\cite{HBR} used by state-of-the-art dynamic race detection frameworks such as \tsan~\cite{Tsan} (section~\ref{sub:tsan-bg}) and \archer~\cite{Archer} (section~\ref{sub:archer-bg}). This section also provides the necessary information on \textit{ARMv8.5-A} hardware-based memory-tagging extension (\mte)~\cite{MTE, MTEWhitepaper}. 

\subsection{Happens-before and Data race}\label{sub:hb-bg}
\textit{Happens-before (HB) relationship}~\cite{HBR} defines partial ordering between events and can be used to infer ordering or lack thereof between any two events of the same or different threads. The partial ordering considers the effects of arbitrary thread interleaving and instruction reordering due to the processor's memory consistency model. 

\begin{align}
	EventType (ET) &= \{RD, WR, LA, LR, S, W\} \\ \notag
	Event (E) &= <ET, TID, MID> 
\end{align}

\textit{\textbf{EventType (ET)}}: An operation type which could be one of read (RD), write (WR), lock acquire(LA) or release (LR), signal (S) or wait (W).

\textit{\textbf{Event}}: An event $E$ is an operation that is represented by a tuple consisting of an $ET$, the thread ID of the issuing thread $TID$, and a unique ID representing the accessed memory $MID$.

\begin{align}
	X_{\tau_{a}} &= E <RD/WR, \tau_{a}, MID > \\ \notag
	Y_{\tau_{b}} &= E <RD/WR, \tau_{b}, MID > 
\end{align}

\begin{align}
	S_{\tau_{a}} &= E <S, \tau_{a}, SID > \\ \notag
	W_{\tau_{b}} &= E <W, \tau_{b}, SID > 
\end{align}

A $HB$ relation between two events $X_{\tau_{a}}$ and $Y_{\tau_{b}}$ issued from thread ID $\tau_{a}$ and $\tau_{b}$ respectively, and accessing the same memory location $MID$ is shown in equation~\ref{eq:HB}. If $X_{\tau_{a}}$ is observed before $Y_{\tau_{b}}$, then $X_{\tau_{a}}$  \textit{happens-before} $Y_{\tau_{b}}$ (represented by $X_{\tau_{a}} \prec Y_{\tau_{b}}$) if either of the following conditions holds: both events are from the same thread, or if the events are from different threads, there exists a transitive event pair $S_{\tau_{a}}$ and $W_{\tau_{b}}$ such that $HB$ can be independently affirmed between each of the pairs $X_{\tau_{a}} \prec S_{\tau_{a}}$, $S_{\tau_{a}} \prec W_{\tau_{b}}$, and $W_{\tau_{b}} \prec Y_{\tau_{b}}$.

\begin{align}\label{eq:HB}
	X_{\tau_{a}} \prec Y_{\tau_{b}} = 
	\begin{cases}
		\tau_{a} = \tau_{b} , \\
		(\tau_{a} \neq \tau_{b}) \land (X_{\tau_{a}} \prec S_{\tau_{a}} \prec W_{\tau_{b}} \prec Y_{\tau_{b}}) 
	\end{cases}
\end{align}

\begin{figure}[tbph]
	\centering
	\includegraphics[width=0.50\linewidth]{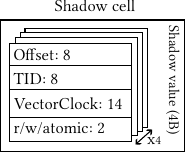}
	\caption{Visual representation of shadow cell used in \tsan and \archer for shared memory access tracking. Every potentially shared 8 byte allocation access is mapped to one of the four shadow values.}
	\label{fig:tsan}
\end{figure}

\paragraph{\textbf{Thread Sanitizer}}\label{sub:tsan-bg}

\tsan~\cite{Tsan} is a well recognized state-of-the-art dynamic race detection framework capable of detecting data races in \textit{Pthread-}based multi-threaded programs. The framework detects data races by tracing $RD$ and $WR$ operations on every potentially shared memory allocation at 8 Bytes (B) granularity. This includes memory allocations in the global or heap segments and those that have references outside a function scope (pointers as formal arguments to functions). The instrumentations necessary to perform the tracing are laid out during the compilation of a target program. During execution, the framework records an event's information $E$ by logging the $ET$, $TID$, and $MID$ for each potentially shared allocation. \tsan uses a 14-bit logical vector clock for $HB$ analysis. The metadata representing the event information is stored as \textit{shadow values} in the program's address space at a fixed offset relative to the address of the memory allocation being accessed. \tsan carves out four such consecutive regions (a \textit{shadow cell}), each of size 4B, to store the metadata for every 8B allocation. Figure~\ref{fig:tsan} provides a visual representation of a shadow cell that consists of four shadow values, each containing the event information of a memory access operation. \tsan also traces lock semantics to identify the set of locks held by a thread at any particular execution point to perform the \textit{lockset} analysis.

\begin{align}\label{eq:tsan-dr}
	Data\:race (X_{\tau_{a}}, Y_{\tau_{b}})  \triangleq & (X_{\tau_{a}} \nprec Y_{\tau_{b}}) \land (Y_{\tau_{b}} \nprec X_{\tau_{a}}) \\ \notag
	& \land (LS(X_{\tau_{a}}) \cap LS(Y_{\tau_{b}}) = \emptyset) \\ \notag
	& \land (IsWrite(X_{\tau_{a}}) \lor IsWrite(Y_{\tau_{b}}))
\end{align}

\tsan defines a data race as an event that satisfied the equation~\ref{eq:tsan-dr}. Two memory access events $X_{\tau_{a}}$ and $Y_{\tau_{b}}$, accessing the exact shared location, are racy if a $HB$ relationship cannot be asserted between them, and the lockset analysis does not identify a common lock between the two events, and one of the events is a $WR$.

\paragraph{\textbf{Archer}}\label{sub:archer-bg}
\archer~\cite{Archer} builds on top of the \tsan framework by more accurately (with lower false positives) detecting data races in \textit{OpenMP-}based multi-threaded applications. It closely models the behavior of \textit{OpenMP} specification~\cite{OpenMP} in its framework to segregate \textit{sequential} code sections of the target application from \textit{may-happen-in parallel} code sections. It then leverages \tsan to only instrument sections of code that could be potentially racy instead of every section potentially accessing shared memory allocation. This focused instrumentation, along with \textit{OpenMP} aware tracing, allows \archer to trace potential data races more accurately in \textit{OpenMP-}based multi-threaded programs.

\begin{figure}[tbph]
	\centering
	\includegraphics[width=1\linewidth]{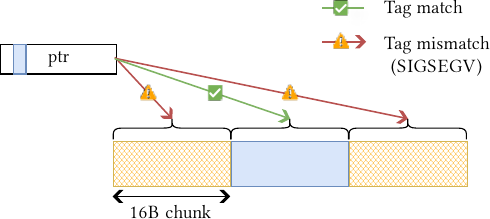}
	\caption{Overview of hardware based memory tagging extension (\mte).}
	\label{fig:mte}
\end{figure}

\subsection{Memory Tagging Extension}\label{sub:mte-bg}
Hardware-based memory tagging extension (\mte)~\cite{MTE, MTEWhitepaper} is an optional feature in chipsets complying with \textit{ARMv8.5-A} architecture or higher. \mte can be used to restrict access of pointers to specified memory granules (16B chunks), resulting in a segmentation fault if a pointer accesses a region outside its stipulated bounds. This access restriction is realized in the hardware by physically tagging 4-bit values (tags) to the accessed memory granule and the pointer used to access the memory. The tags are part of the top-byte range of the pointer, whereas the tags applied to memory granules are architecture-specific and hidden from userspace access. Figure~\ref{fig:mte} provides an overview of \mte. A pointer \textit{ptr}, which has been tagged with a value "\textit{blue}" is restricted from accessing memory granules that are tagged "\textit{yellow}." The processor catches access to a memory granule outside its matching tag range and raises it as a segmentation fault. A point to note is that the \mte processor extension operates at 16B granularity: the smallest memory unit that can be tagged and checked by the processor for a tag mismatch is a granule. The tag size of \textit{4 bits} restricts the tag range between 0 and 15 ($tag = [0, 15]$). The processor also requires taggable memory granules to be 16B aligned. The processor can be instructed to perform tag-based operations in \mte enabled userspace process using \textit{STG} to set the granule tag, \textit{IRG} to randomize and update the pointer tag, \textit{SUBG} to decrement pointer tag, and \textit{ADDG} to increment pointer tag.
\section{Requirements and Scope}\label{sec:assumptions}

With the intent to build a dynamic data race detection framework capable of running in a production environment, we identify the following requirements for \sys:

\paragraph{\textbf{Requirement-1: High precision and accuracy}} Dynamic race detection frameworks should be capable of detecting and reporting race with higher accuracy: reporting ground truth of the application's memory access states, and higher precision: report \textit{true} races deterministically on every execution. We enforce these statistical requirements on \sys even during \textit{non-overlapping-but-observed} racy interleaving: the framework should report true races based on the history of observed events. \sys should report true races with high confidence, even on a single execution of an instrumented target application. 

\paragraph{\textbf{Requirement-2: No false positive}} Typically, fixing a data race requires considerable debugging effort from a developer. If a race detection framework incurs false positives, the developer must spend additional effort validating the reported race. To avoid this burden on the developer, \sys has to report no false positives.

\paragraph{\textbf{Requirement-3: Low processor and memory overhead}}
Most data races are fixed during the development and testing phases in an application's life cycle.
However, as a corollary of Murphy's law~\cite{Murphy}, \textit{hard-to-detect} data races could exist hidden in multiple releases until finally observed in a production environment due to atypical racy interleavings. 
%
%
\sys is intended to work in a production environment and should be designed to leverage hardware-based primitives to reduce the overhead of code instrumentation for data race detection.

\paragraph{\textbf{Requirement-4: On-the-fly data race reporting}} ~\sys should be designed to synchronously report data races when a racy memory access is performed. This allows for accurate causal analysis to root cause a data race quickly.

\paragraph{\textbf{Requirement-5: System-wide memory-tag reusability}} The usage of hardware-based memory-tagging in \sys instrumented target applications should not devoid use of memory-tagging for out-of-bounds or use-after-free access detection by other processes of the system or by the kernel.

\paragraph{\textbf{Scope}} While hardware-based memory-tagging for race detection is applicable to both procedural and object-oriented programming languages, \sys prototypes the framework for native C applications using a shared-memory programming paradigm. \sys can detect data races in the global, heap, and stack segments for \textit{Pthread-} and \textit{OpenMP-}based multi-threaded applications. \sys does not separately instrument shared libraries but provides the option to statically link shared libraries to detect potential data races between shared libraries and the application. The current prototype of \sys does not support data races arising from target offloads and vectorization instructions; we consider these as part of future extensions to \sys. 
\section{Design and Implementation}\label{sec:design}

\begin{figure}[tbph]
	\centering
	\includegraphics[width=1\linewidth]{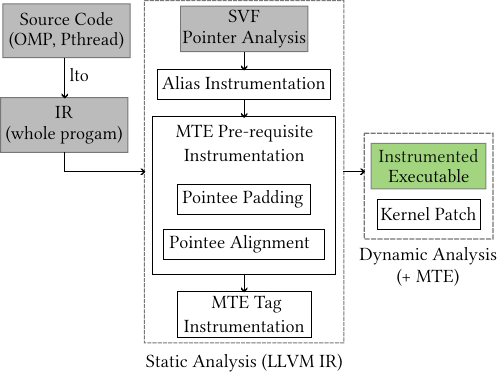}
	\caption{Overview of \sys framework. Blocks highlighted in \textit{gray} signify unmodified modules.}
	\label{fig:mtcsan-fw}
\end{figure}

Figure~\ref{fig:mtcsan-fw} provides an overview of \sys framework. The analysis and instrumentation passes of the framework are implemented on \textit{llvm-16.0.6}~\cite{LLVM:CGO04}. The passes are common for both OpenMP and Pthread applications since, during the compiler's intermediate stage, OpenMP degrades to Pthread-based intermediate representation (OpenMP is implemented using Pthread libraries). \mte~\cite{MTE, MTEWhitepaper} in Linux kernel requires setting \textit{PROT\_MTE}~\cite{MTERequirements} page attribute on anonymous pages targeted by the memory-tagging instructions. We chose to implement the necessary changes on \textit{filesystem (fs)} and \textit{memory management (mm)} modules of the Linux kernel as the kernel sets up the stack and global segments before invoking the dynamic loader.

\subsection{Definitions}\label{sub:def}

This section defines the necessary terms for formally defining the core algorithm of \sys and its implementation. We, furthermore, redefine the terms \textit{event, event type, and data race} by building on the \tsan's~\cite{Tsan} formalization.

\paragraph{Pointee ${P}^{m}$} Pointee is a 16B aligned shared allocation defined as a simple data, array, or user-defined type. It is aligned and padded to be a memory allocation of a multiple of 16B, as described in~\ref{sub:req}.

\paragraph{Memory granules} It represents a 16B aligned chunk of memory that can be memory tagged. 

\paragraph{Pointee granule ${P}^{m}_{x}$} It represents a granule with an index $x$ into a shared pointee $P^{m}$.

\paragraph{Thread-ID $\tau_{a}$} It represents a monotonically increasing thread ID obtained by subtracting the unique TID~\cite{TID} of the main thread from the TID of the thread under consideration.

\paragraph{Pointer $p_{m,i, \tau_{a}}$} It represents a thread-local variable $i$ of a thread $\tau_{a}$ pointing to the pointee granule ${P}^{m}_{x}$.

\begin{align}
	Tag (T) &= \{t | t \in \mathbb{W},  0 \leq t < 16\}  \\ \notag
	EventType (ET) &= \{RD, WR\} \\
	Event (E) &= <ET, TID, MID, T>
\end{align}

\paragraph{Memory access event type (ET)} An operation type which could be either a read $RD$ or a write $WR$ operation to a shared pointee granule ${P}^{m}_{x}$.

\paragraph{Memory access events (E)} An event $E$ is an operation that is represented by a tuple consisting of an $ET$, thread ID of the issuing thread $TID$, address of the accessed memory $MID$, and a memory tag $T$.  

\paragraph{Segments $S_{m, i, \tau_{a}}$} It represents a bounded region consisting of events, during the execution, that associates an access attribute of a pointee granule ${P}^{m}_{x}$ to a pointer $p_{m, i,  \tau_{a}}$. The access attribute, which can be either \textit{Exclusive} or \textit{Shared}, is set at a segment's first memory access event for a particular granule of a thread. The pointer retains the access attribute until either the thread terminates or until another valid (non-racy) pointer accesses ${P}^{m}_{x}$. On thread termination, the exclusive access of ${P}^{m}_{x}$ is passed onto its parent thread.

\paragraph{Lockset state $LSS_{{P}^{m}}$} The lockset $P^{m}_{meta}.LS$ of a pointee ${P}^{m}$ represents the set of locks (\textit{mutually exclusive} or \textit{reader-shared}) taken by threads in the past to access any granule of ${P}^{m}$. The lockset state $LSS_{{P}^{m}}$ is a tri-state value representing the result of lockset analysis with values that are \textit{inconclusive}, conclusive with mutually \textit{exclusive} lock usage, or conclusive with \textit{shared} reader lock usage (e.g., RW locks). Inconclusive lockset analysis resulting from incorrect or missing locks is insufficient to assert data race, as pointers can read or write to a shared allocation at different granules without resulting in concurrency faults.

\begin{align}
	X_{\tau_{a}} &= E <RD/WR, \tau_{a}, {P}^{m}_{x}, T_{x} > \\ \notag
	Y_{\tau_{b}} &= E <RD/WR, \tau_{b}, {P}^{m}_{x}, T_{y}> 
\end{align}

\begin{align}
	TagCheck(E, MID) = 
	\begin{cases}
		0, & \text{if } Tag(E) \neq Tag(MID) \\
		1, &  \text{if } Tag(E) = Tag(MID) \\
	\end{cases}
\end{align}

\begin{align}\label{eq:mtcsan-dr}
	Data\:race (X_{\tau_{a}}, Y_{\tau_{b}})  \triangleq & (TagCheck(X_{\tau_{a}},  {P}^{m}_{x}) \land \\ \notag& \neg TagCheck(Y_{\tau_{b}},  {P}^{m}_{x})) \\ \notag
	& \land (LS(X_{\tau_{a}}) \cap LS(Y_{\tau_{b}}) = \emptyset) \\ \notag
	& \land (IsWrite(X_{\tau_{a}}) \lor IsWrite(Y_{\tau_{b}}))
\end{align}

\paragraph{Data race} \sys defines data race as a condition which satisfies equation~\ref{eq:mtcsan-dr}. Two memory access events, $X_{\tau_{a}}$ and $Y_{\tau_{b}}$ accessing the same shared pointee granule ${P}^{m}_{x}$ such that $X_{\tau_{a}}$ is observed before $Y_{\tau_{b}}$, are racy if a tag check $TagCheck(E, MID)$ (checked through \mte processor intrinsics) on $X_{\tau_{a}}$ succeeds but fails on $Y_{\tau_{b}}$, and if the \textit{lockset} analysis on the two events does not yield a common lock (inconclusive), and if either of the event is a $WR$ operation. 

\paragraph{Reader ILU}~\label{par:ILU} If both the events are $RD$, and if the tag check on $Y_{\tau_{b}}$ fails, with the \textit{lockset} analysis yielding no common between the events, then \sys categorizes such instances as \textit{inconsistent lock usage}.

\begin{algorithm}[tbph]
	\caption{Lockset analysis algorithm.}\label{alg:lock}
	\begin{algorithmic}[1]
		\State \textbf{handle\_lock\_event}($L_{p}$) \label{subalgo: lock}
		\LeftComment{Update $\tau_{a}$ lockset.}
		\State \quad $LS_{\tau_{a}} \leftarrow LS_{\tau_{a}} \cup \{L_{p}\} $
		\newline
		
		\State \textbf{handle\_unlock\_event}($L_{p}$) \label{subalgo: unlock}
		\LeftComment{Update $\tau_{a}$ lockset.}
		\State \quad $LS_{\tau_{a}} \leftarrow LS_{\tau_{a}} - \{L_{p}\} $
		\State \quad \textbf{update\_tag\_pointer}($p_{m,i, \tau_{a}}$, 0) \Comment{$\forall$ updated pointers in the segment.}
		\newline
		
		\State \textbf{update\_lockset\_pointee}($LS_{\tau_{a}}, P^{m}_{meta}.LS$) \label{subalgo: ls_update}
		\LeftComment{Update or trim existing pointee lockset $P^{m}_{meta}.LS$.}
		\If{$P^{m}_{meta}.LS == \emptyset$}
		\If{$LS_{\tau_{a}} == \emptyset$}
		 \LeftComment{\quad Accessed before with reader-shared lock or no lock.}
		\State $P^{m}_{meta}.SPA \leftarrow true$
		\Else
		\State $P^{m}_{meta}.SPA \leftarrow false$ 
		\State $P^{m}_{meta}.LS \leftarrow LS_{\tau_{a}}$
		\EndIf
		\Else
		\State $P^{m}_{meta}.LS \leftarrow P^{m}_{meta}.LS \cap LS_{\tau_{a}}$ \Comment{Trim the lockset.}
		\EndIf
		\newline
		
		\State \textbf{handle\_lockset\_check}($LS_{\tau_{a}}$, ${P}^{m}_{x}$) \label{subalgo: ls_check}
		\LeftComment{Update $P^{m}_{meta}$ lockset.}
		\LeftComment{Return: }
		\LeftComment{\quad \quad \quad \textit{Inconclusive} if lockset fails.}
		\LeftComment{\quad \quad \quad \textit{Exclusive} if lockset passes and no reader-shared locks.}
		\LeftComment{\quad \quad \quad \textit{Shared} if lockset passes and at least one reader-shared lock.}
		\State $P^{m}_{meta}$ $\leftarrow$ \textbf{get\_pointee\_metadata}(${P}^{m}_{x}$)
		\State $P^{m}_{meta}.SPA$ $\leftarrow$ \textbf{is\_prioraccess\_shared}(${P}^{m}_{x}$)
		\If {{$P^{m}_{meta}.LS$ == $\emptyset$} \&\&  {$P^{m}_{meta}.SPA$}} 
		\State return Inconclusive \Comment{Maybe-racy}
		\ElsIf{{$P^{m}_{meta}.LS$ == $\emptyset$} || {$LS_{\tau_{a}} \cap P^{m}_{meta}.LS$}}
		\State $\mathbf{P^{m}_{meta}.MU.Lock()}$ \Comment{Atomically update $P^{m}_{meta}.LS$.}
		\State  $P^{m}_{meta}.LS$ $\leftarrow$ \textbf{update\_lockset\_pointee}($LS_{\tau_{a}}, P^{m}_{meta}.LS$)
		\State $\mathbf{P^{m}_{meta}.MU.Unlock()}$
		\If {\textbf{has\_rshared\_locks}($LS_{\tau_{a}} \cap P^{m}_{meta}.LS$)}
		\State return Shared \Comment{Non-racy}
		\Else
		\State return Exclusive \Comment{Non-racy}
		\EndIf
		\Else
		\State return Inconclusive \Comment{Maybe-racy}
		\EndIf
	\end{algorithmic}
\end{algorithm}

\begin{algorithm}[tbph]
	\caption{Read/Write instrumentation for segment $S_{m, i, \tau_{a}}$, and granule re-target.}\label{alg:seg}
	\begin{algorithmic}[1]
		\State \textbf{handle\_read\_write\_segment}($LS_{\tau_{a}}$, $p_{m,i, \tau_{a}}$, \textit{IsRead}) \label{subalgo: read_write}
		\LeftComment{Instrument tag update for each read or write segment}
		\State \quad ${P}^{m}_{x}$ $\leftarrow$ \textbf{get\_pointee}($p_{m,i, \tau_{a}}$)
		\State \quad $LSS_{{P}^{m}}$ $\leftarrow$ \textbf{handle\_lockset\_check}($ LS_{\tau_{a}}$, ${P}^{m}_{x}$)
		\If {$LSS_{{P}^{m}}$ == \textit{Inconclusive}} \Comment{Maybe-racy}
		\State \textbf{insert\_dummy\_load}($p_{m,i, \tau_{a}}$) \Comment{$\forall$ granule re-target.}  \label{subalgo: dummy_load}
		\ElsIf{$LSS_{{P}^{m}}$ == \textit{Exclusive}} \Comment{Non-racy}
		\LeftComment{$L_p \in P^{m}_{meta}.LS$}
		\LeftComment{$P^{m}_{meta}.LS$ == $\emptyset$}
		\ElsIf{$LSS_{{P}^{m}}$ == \textit{Shared}} \Comment{Non-racy}
		\LeftComment{$P^{m}_{meta}.LS$ contains reader-shared locks.}
		\If{\textit{IsRead}}
		\State $tag \leftarrow$ \textbf{get\_tag}(${P}^{m}_{x}$)\label{subalgo: tag_inherit}
		\State $Skip\_tag\_pointee \leftarrow true$
		\State \textbf{update\_tag\_pointer}($p_{m,i, \tau_{a}}$, $tag$) \Comment{$\forall$ granule re-target.}
		\EndIf
		\EndIf
		\State \textbf{update\_tag\_pointer}($p_{m,i, \tau_{a}}$, $\tau_{a}$) \Comment{$\forall$ granule re-target.} \label{subalgo: tag_update}
		\If{\textbf{not} $Skip\_tag\_pointee$}
		\State \textbf{update\_tag\_pointee}($p_{m,i, \tau_{a}}$) \Comment{$\forall$ granule re-target.}
		\EndIf
	\end{algorithmic}
\end{algorithm}

\subsection{TBRI: Core algorithm}\label{sub:algo}

\sys combines lockset analysis (algo~\ref{alg:lock}) with structured \mte tag instrumentation (algo~\ref{alg:seg}) to detect data races. \sys terms this approach \textit{TBRI: Tag-based race inference}. The framework detects racy memory access events through the processor's reported memory-tag mismatches. 

\paragraph{Lockset analysis} \sys maintains two separate lockset information: per thread lockset $LS_{\tau_{a}}$, and per pointee lockset $P^{m}_{meta}.LS$.  $LS_{\tau_{a}}$ tracks currently held locks by a thread at all execution points, with the lockset updated when a lock (algo~\ref{alg:lock}:~\ref{subalgo: lock}) or unlock (algo~\ref{alg:lock}:~\ref{subalgo: unlock}) event is executed by the thread. $P^{m}_{meta}.LS$ tracks the minimal lockset with which any thread executes memory access events to a particular shared pointe $P^{m}$. This lockset is iteratively trimmed with each segment access to produce a precise and minimal set of locks that need to be held by any thread to perform subsequent memory accesses on a pointee granule ${P}^{m}_{x}$.  (algo~\ref{alg:lock}:~\ref{subalgo: ls_update}).

\sys then selectively enables the segment and granule-retarget instrumentations during execution time (algo~\ref{alg:seg}) based on the lockset state $LSS_{{P}^{m}}$ obtained during segment setup (algo~\ref{alg:lock}:~\ref{subalgo: ls_check}). $LSS_{{P}^{m}}$ reports \textit{Inconclusive} if $LS_{\tau_{a}}$ and $P^{m}_{meta}.LS$ do not have a common lock between them. Suppose the reported result is conclusive due to at least one common lock between the locksets or when $P^{m}_{meta}.LS$ is empty (from initial read or write), the result is reported as \textit{Shared} if there is at least one shared lock event or \textit{Exclusive} otherwise. \textit{Shared} lockset state allows for inheriting tag from ${P}^{m}_{x} $ to allow \textit{READ/ READ} events when using reader-shared locks.

\paragraph{Segment and granule-retarget} \sys selectively enables the required tag instrumentation based on lockset states $LSS_{{P}^{m}}$ computed during segment setup. 

If $LSS_{{P}^{m}}$ is \textit{Inconclusive}, the framework triggers execution of a dummy load instruction using un-tagged pointer $p_{m, i, \tau_{a}}$ (algo~\ref{alg:seg}:~\ref{subalgo: dummy_load}). If the pointee granule ${P}^{m}_{x}$ was previously tagged, then this dummy load instruction results in a memory-tag mismatch, inferring a data race. If the processor did not report a tag mismatch on the dummy load, then \sys triggers execution of tag update instructions for both $p_{m, i, \tau_{a}}$ and ${P}^{m}_{x}$ (algo~\ref{alg:seg}:~\ref{subalgo: tag_update}).

If $LSS_{{P}^{m}}$ is \textit{Shared} or \textit{Exclusive}, then \sys triggers execution of tag update instructions for $p_{m, i, \tau_{a}}$ and ${P}^{m}_{x}$ along with an additional instrumentation to inherit tag (algo~\ref{alg:seg}:~\ref{subalgo: tag_inherit}) if \textit{Shared} and when the memory event is \textit{READ}.

\subsection{Implementation}\label{sub:impl}

\paragraph{Pointer analysis} \sys implements a context-sensitive interprocedural pointer analysis by extending the SVF's interprocedural points-to analysis~\cite{SVF}. The pointer analysis pass reduces instrumentation scope (thereby reducing instrumentation overhead) by identifying memory events that perform access to shared allocations. It builds an association between all pointer aliases $p_{m, i}$ to a shared pointee ${P}^{m}$. The resulting alias set $AS: {{P}^{m}}, [p_{m, 1}, p_{m, 2}.. p_{m, i}]$ represents a many-to-many relationship between pointers and pointees of the set, with the pointer unrestrained to target any granule of the shared pointee or a different shared pointee altogether. The granule targeted by the pointer is identified by tracing the \textit{GetElementPtr}~\cite{GEP} instruction during the analysis pass. The performed pointer analysis enforces soundness (does not miss identifying aliases to shared pointee) by over-approximating the alias set. The subsequent stages of the framework consume \textit{AS} to perform the necessary analysis and instrumentation for detecting data race.

\paragraph{Alias instrumentation} Pointers $p_{m, i}$ of alias set $AS$ can be shared across multiple threads (\textit{non thread-private}) if passed around as formal parameters to functions consuming it. Detecting data race through memory-tagging requires the creation of \textit{thread-private pointer aliases} $p_{m, i, \tau_{a}}$ for each $p_{m, i}$ of the $AS$. $p_{m, i, \tau_{a}}$ is necessary since, without such a thread-private variable, a tag updated by a thread in a non-thread-private pointer can propagate to another, resulting in missed race detection opportunities. $p_{m,i, \tau_{a}}$  is set to point to the same shared pointee granule ${P}^{m}_{x}$. It is, however, maintained private per thread $\tau_{a}$, thereby avoiding unintended tag propagation. The address-taken value of $p_{m, i, \tau_{a}}$  is updated to shadow their respective pointer's $p_{m, i}$ value within each segment. 

\paragraph{MTE setup}\label{sub:req} In order to facilitate memory-tagged access through \mte, every contiguous shared pointee $P_{m}$ in the alias set $AS$ needs to be transformed into 16B aligned allocation ${P_{m}}$ which is also a multiple of 16B. The padding and alignment process increases the overall size of the ${P_{m}}$ to a multiple of 16B and aligns the base of the memory allocation to its natural size boundary. For example, if we consider an array of integer type to be a shared allocation with 64 entries (\texttt{int [64]}), then the resulting shared allocation would be modified to an array of struct type \{\texttt{int, int}\} with the same number of indices as the original array. The framework requires transforming every shareable pointee, identified by the pointer analysis pass, to a multiple of 16B by considering the granular unit within the shared pointee. We discuss a potential approach to alleviate this memory overhead in section~\ref{sec:limitations}. This pass also enables the application to use synchronous memory-tagging.

\paragraph{Race instrumentation}  This pass performs the core instrumentation that enforces access constraints on shared pointees ${P}^{m}_{x}$. The access is enforced through structured \mte intrinsic instrumentation in segments which reference ${P}^{m}_{x}$. Algorithms~\ref{alg:lock}, ~\ref{alg:seg} along with section~\ref{sub:algo} describe the lockset and segment instrumentation performed by this pass. The algorithm mimics \textit{happens-before} analysis by using memory tags to discern previous memory access events, with lockset analysis providing the runtime condition to enable or disable tag updates. If the lockset analysis $LSS_{{P}^m}$ is conclusive (\textit{Shared} or \textit{Exclusive}), then the memory-tag of both the pointer $p_{m, i, \tau_{a}}$ and pointee granule ${P}^{m}_{x}$ are updated for future data race checks. If $LSS_{{P}^m}$ is \textit{Inconclusive}, then a dummy load instruction is executed during the process's execution time to detect potential data race before updating the tags. The lockset checks (algo~\ref{alg:lock}:~\ref{subalgo: ls_check}) are performed only during segment setup (which can span multiple functions) with the tag updates (algo~\ref{alg:seg}:~\ref{subalgo: tag_update}) performed on every granule-retarget (when $p_{m, i, \tau_{a}}$ is updated to point to a different granule).

\paragraph{Race signal handler} \sys leverages hardware-based memory tagging to detect tag mismatches resulting from data race. The processor synchronously reports tag mismatch and triggers a registered signal handler. This pass sets up the post-mortem sigaction, which reports the data race on detecting \textit{SIGSEGV} from \mte. The sigaction prints the faulting instruction and the tag information to identify the competing threads in a data race. The sigaction can be used with a debugger to perform precise post-mortem analysis.

\begin{figure*}[tbph]
	\centering
	\includegraphics[width=0.90\textwidth]{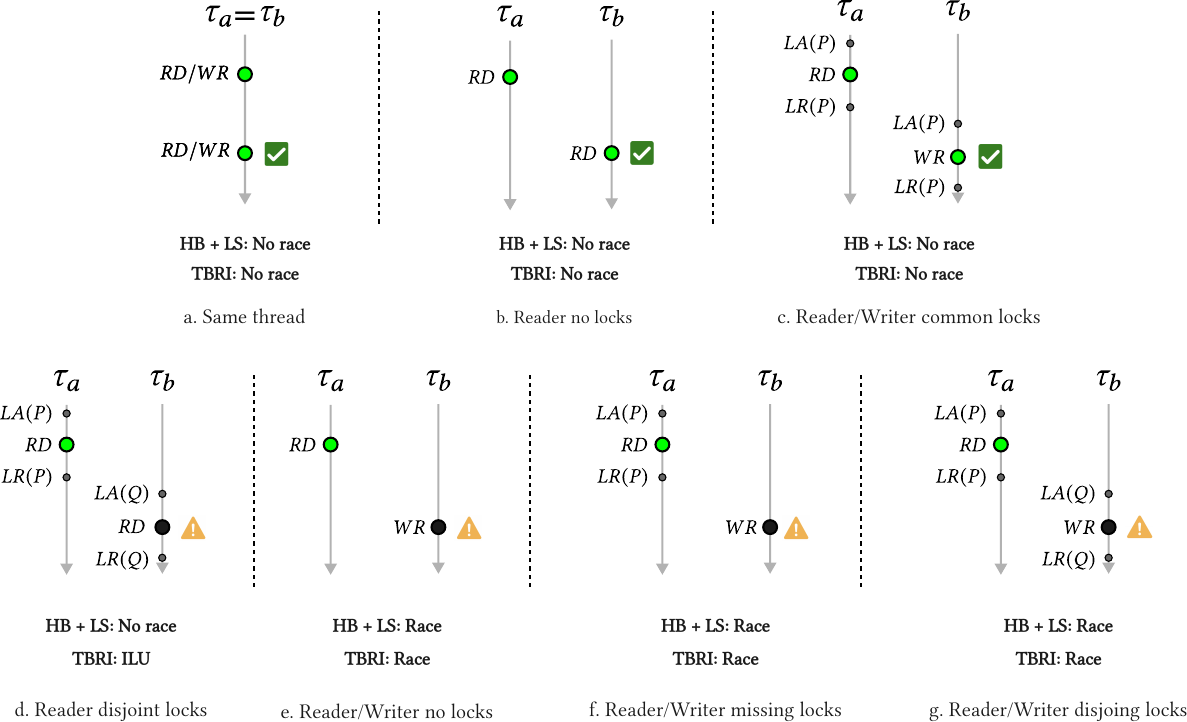}
	\caption{Comparison of \textit{HB + LS} with \textit{TBRI}. A check signifies \mte tag match (no race), and an exclamation signifies \mte tag mismatch (race). $RD/WR$: Read/write events to the same pointee granule. $LA(x)/LR(x)$: Lock acquision/release for lock $x$. $\tau_{a}, \tau_{b}$: Thread IDs representing two threads.}
		\label{fig:tbri}
\end{figure*}

\subsection{Race detection cases}\label{sub:cases}
Figure~\ref{fig:tbri} showcases a few combinations of non-overlapping memory access events $RD/WR$ accessing the same shared memory granule. The cases compare inference results from \textit{happens-before (HB)}~\cite{HBR} and \textit{lockset}~\cite{LocksetAnalysis} analysis with that of \textit{TBRI} algorithm used by \sys. Cases \textit{a-c} show non-racy events between threads $\tau_{a}$ and $\tau_{b}$. Both \textit{HB+LS} and \textit{TBRI} report correct \textit{no-race} results in these cases. Cases \textit{e-f} show racy events with no synchronization (case-e), missing lock access (case-f), and inconsistent synchronization (case-g). In each of these cases, both the analysis approaches report data races, with \sys using the tag mismatch to assert the same. Case-d shows an ILU scenario (section~\ref{par:ILU}), where two $RD$ access events have disjoint locksets. \sys flags these cases as they imply semantic anomalies and could potentially cause subsequent data races.

\section{Evaluation and Discussion}\label{sec:eval}

This section addresses the following questions regarding the effectiveness of the \sys framework:

\begin{itemize}
	\item How accurate and precise is \sys when detecting data race in \textit{Pthread-} and \textit{OpenMP-}based applications? Moreover, how does this compare against current state-of-the-art dynamic race detection frameworks? (Section~\ref{sub:precision})
	\item In which scenarios do \sys perform better or worse than current state-of-the-art dynamic race detection frameworks? (Section~\ref{sub:precision})
	\item What is the execution and memory overhead of the instrumentations carried out in \sys? Moreover, how does this compare against current state-of-the-art dynamic race detection frameworks? (Section~\ref{sub:execution-overhead})
	\item How does \sys perform with variations in reader/writer thread counts, number of shared allocations, and lockset acquisitions?(Section~\ref{sub:causation})
	\item Has the framework successfully detected new races in large-scale applications? (Section~\ref{sub:new-race})
\end{itemize}

\paragraph{\textbf{Evaluation setup}}
We compare
\sys against \tsan~\cite{Tsan} and \archer~\cite{Archer}, current state-of-the-art happens-before and lockset-based dynamic data race detection frameworks. Both are compiled from LLVM (\textit{llvmorg-16.0.6})~\cite{LLVM:CGO04} compiler toolchain. The target programs using these frameworks are compiled with the \texttt{-fsanitize=thread} flag and \texttt{-O2} optimizations. As \archer runtime is enabled by default, benchmarks evaluating performance and accuracy for \tsan are set up with the environment variable \\ \texttt{ARCHER\_OPTIONS="enabled=0"}.

We use \textit{Pixel 8 (Shiba)}~\cite{Pixel8} android smartphones for our evaluation as the \textit{Tensor G3} chipset provides \mte processor intrinsic. The smartphone contains eight general-purpose CPUs running on \textit{GrapheneOS-14}~\cite{GrapheneOS} with patched \textit{5.15.148-android14} Linux kernel. We evaluate our benchmarks on the \textit{Ubuntu 22.04} distribution using \textit{UserLand}~\cite{Uland}. All cores of the smartphone are locked at 1.7GHz using \textit{CPUFreq governor}~\cite{GOV} set to \textit{performance} mode.

\begin{table*}[tbph]
	\centering
	\resizebox{\textwidth}{!}{%
\begin{tabular}{|ccccccccccccccccccccccccc|}
	\hline
	\multicolumn{25}{|c|}{\textbf{DRB 1.4.1 (208)}} \\ \hline
	\multicolumn{1}{|c|}{\multirow{3}{*}{\textbf{\begin{tabular}[c]{@{}c@{}}Benchmark Category \\ (\#Microbenchmarks/ \#TP)\end{tabular}}}} &
	\multicolumn{8}{c|}{\textbf{\tsan (llvmorg-16.0.6)}} &
	\multicolumn{8}{c|}{\textbf{\archer (llvmorg-16.0.6)}} &
	\multicolumn{8}{c|}{\textbf{\sys}} \\ \cline{2-25} 
	\multicolumn{1}{|c|}{} &
	\multicolumn{1}{c|}{\multirow{2}{*}{\textbf{FP}}} &
	\multicolumn{1}{c|}{\multirow{2}{*}{\textbf{TN}}} &
	\multicolumn{3}{c|}{\textbf{\begin{tabular}[c]{@{}c@{}}FN \\ (\#exec)\end{tabular}}} &
	\multicolumn{3}{c|}{\textbf{\begin{tabular}[c]{@{}c@{}}TP \\ (\#exec)\end{tabular}}} &
	\multicolumn{1}{c|}{\multirow{2}{*}{\textbf{FP}}} &
	\multicolumn{1}{c|}{\multirow{2}{*}{\textbf{TN}}} &
	\multicolumn{3}{c|}{\textbf{\begin{tabular}[c]{@{}c@{}}FN\\ (\#exec)\end{tabular}}} &
	\multicolumn{3}{c|}{\textbf{\begin{tabular}[c]{@{}c@{}}TP\\ (\#exec)\end{tabular}}} &
	\multicolumn{1}{c|}{\multirow{2}{*}{\textbf{FP}}} &
	\multicolumn{1}{c|}{\multirow{2}{*}{\textbf{TN}}} &
	\multicolumn{3}{c|}{\textbf{\begin{tabular}[c]{@{}c@{}}FN\\ (\#exec)\end{tabular}}} &
	\multicolumn{3}{c|}{\textbf{\begin{tabular}[c]{@{}c@{}}TP\\ (\#exec)\end{tabular}}} \\ \cline{4-9} \cline{12-17} \cline{20-25} 
	\multicolumn{1}{|c|}{} &
	\multicolumn{1}{c|}{} &
	\multicolumn{1}{c|}{} &
	\multicolumn{1}{c|}{\textbf{\#1}} &
	\multicolumn{1}{c|}{\textbf{\#5}} &
	\multicolumn{1}{c|}{\textbf{\#20}} &
	\multicolumn{1}{c|}{\textbf{\#1}} &
	\multicolumn{1}{c|}{\textbf{\#5}} &
	\multicolumn{1}{c|}{\textbf{\#20}} &
	\multicolumn{1}{c|}{} &
	\multicolumn{1}{c|}{} &
	\multicolumn{1}{c|}{\textbf{\#1}} &
	\multicolumn{1}{c|}{\textbf{\#5}} &
	\multicolumn{1}{c|}{\textbf{\#20}} &
	\multicolumn{1}{c|}{\textbf{\#1}} &
	\multicolumn{1}{c|}{\textbf{\#5}} &
	\multicolumn{1}{c|}{\textbf{\#20}} &
	\multicolumn{1}{c|}{} &
	\multicolumn{1}{c|}{} &
	\multicolumn{1}{c|}{\textbf{\#1}} &
	\multicolumn{1}{c|}{\textbf{\#5}} &
	\multicolumn{1}{c|}{\textbf{\#20}} &
	\multicolumn{1}{c|}{\textbf{\#1}} &
	\multicolumn{1}{c|}{\textbf{\#5}} &
	\textbf{\#20} \\ \hline
	\multicolumn{1}{|c|}{\textbf{LD (86/ 40)}} &
	\multicolumn{1}{c|}{23} &
	\multicolumn{1}{c|}{23} &
	\multicolumn{1}{c|}{4} &
	\multicolumn{1}{c|}{2} &
	\multicolumn{1}{c|}{2} &
	\multicolumn{1}{c|}{36} &
	\multicolumn{1}{c|}{38} &
	\multicolumn{1}{c|}{38} &
	\multicolumn{1}{c|}{2} &
	\multicolumn{1}{c|}{44} &
	\multicolumn{1}{c|}{6} &
	\multicolumn{1}{c|}{6} &
	\multicolumn{1}{c|}{6} &
	\multicolumn{1}{c|}{34} &
	\multicolumn{1}{c|}{34} &
	\multicolumn{1}{c|}{34} &
	\multicolumn{1}{c|}{0} &
	\multicolumn{1}{c|}{46} &
	\multicolumn{1}{c|}{5} &
	\multicolumn{1}{c|}{4} &
	\multicolumn{1}{c|}{4} &
	\multicolumn{1}{c|}{35} &
	\multicolumn{1}{c|}{36} &
	36 \\ \hline
	\multicolumn{1}{|c|}{\textbf{SYN (36/ 19)}} &
	\multicolumn{1}{c|}{13} &
	\multicolumn{1}{c|}{4} &
	\multicolumn{1}{c|}{3} &
	\multicolumn{1}{c|}{2} &
	\multicolumn{1}{c|}{2} &
	\multicolumn{1}{c|}{17} &
	\multicolumn{1}{c|}{17} &
	\multicolumn{1}{c|}{17} &
	\multicolumn{1}{c|}{0} &
	\multicolumn{1}{c|}{17} &
	\multicolumn{1}{c|}{3} &
	\multicolumn{1}{c|}{3} &
	\multicolumn{1}{c|}{3} &
	\multicolumn{1}{c|}{16} &
	\multicolumn{1}{c|}{16} &
	\multicolumn{1}{c|}{16} &
	\multicolumn{1}{c|}{0} &
	\multicolumn{1}{c|}{17} &
	\multicolumn{1}{c|}{4} &
	\multicolumn{1}{c|}{4} &
	\multicolumn{1}{c|}{4} &
	\multicolumn{1}{c|}{15} &
	\multicolumn{1}{c|}{15} &
	15 \\ \hline
	\multicolumn{1}{|c|}{\textbf{EB (8/ 3)}} &
	\multicolumn{1}{c|}{5} &
	\multicolumn{1}{c|}{0} &
	\multicolumn{1}{c|}{0} &
	\multicolumn{1}{c|}{0} &
	\multicolumn{1}{c|}{0} &
	\multicolumn{1}{c|}{3} &
	\multicolumn{1}{c|}{3} &
	\multicolumn{1}{c|}{3} &
	\multicolumn{1}{c|}{0} &
	\multicolumn{1}{c|}{5} &
	\multicolumn{1}{c|}{2} &
	\multicolumn{1}{c|}{2} &
	\multicolumn{1}{c|}{2} &
	\multicolumn{1}{c|}{1} &
	\multicolumn{1}{c|}{1} &
	\multicolumn{1}{c|}{1} &
	\multicolumn{1}{c|}{0} &
	\multicolumn{1}{c|}{5} &
	\multicolumn{1}{c|}{1} &
	\multicolumn{1}{c|}{1} &
	\multicolumn{1}{c|}{1} &
	\multicolumn{1}{c|}{2} &
	\multicolumn{1}{c|}{2} &
	2 \\ \hline
	\multicolumn{1}{|c|}{\textbf{OOB (2/ 2)}} &
	\multicolumn{1}{c|}{0} &
	\multicolumn{1}{c|}{0} &
	\multicolumn{1}{c|}{0} &
	\multicolumn{1}{c|}{0} &
	\multicolumn{1}{c|}{0} &
	\multicolumn{1}{c|}{2} &
	\multicolumn{1}{c|}{2} &
	\multicolumn{1}{c|}{2} &
	\multicolumn{1}{c|}{0} &
	\multicolumn{1}{c|}{0} &
	\multicolumn{1}{c|}{0} &
	\multicolumn{1}{c|}{0} &
	\multicolumn{1}{c|}{0} &
	\multicolumn{1}{c|}{2} &
	\multicolumn{1}{c|}{2} &
	\multicolumn{1}{c|}{2} &
	\multicolumn{1}{c|}{0} &
	\multicolumn{1}{c|}{0} &
	\multicolumn{1}{c|}{0} &
	\multicolumn{1}{c|}{0} &
	\multicolumn{1}{c|}{0} &
	\multicolumn{1}{c|}{2} &
	\multicolumn{1}{c|}{2} &
	2 \\ \hline
	\multicolumn{1}{|c|}{\textbf{SH (3/ 3)}} &
	\multicolumn{1}{c|}{0} &
	\multicolumn{1}{c|}{0} &
	\multicolumn{1}{c|}{1} &
	\multicolumn{1}{c|}{1} &
	\multicolumn{1}{c|}{1} &
	\multicolumn{1}{c|}{2} &
	\multicolumn{1}{c|}{2} &
	\multicolumn{1}{c|}{2} &
	\multicolumn{1}{c|}{0} &
	\multicolumn{1}{c|}{0} &
	\multicolumn{1}{c|}{1} &
	\multicolumn{1}{c|}{1} &
	\multicolumn{1}{c|}{1} &
	\multicolumn{1}{c|}{2} &
	\multicolumn{1}{c|}{2} &
	\multicolumn{1}{c|}{2} &
	\multicolumn{1}{c|}{0} &
	\multicolumn{1}{c|}{0} &
	\multicolumn{1}{c|}{0} &
	\multicolumn{1}{c|}{0} &
	\multicolumn{1}{c|}{0} &
	\multicolumn{1}{c|}{3} &
	\multicolumn{1}{c|}{3} &
	3 \\ \hline
	\multicolumn{1}{|c|}{\textbf{TS (31/ 15)}} &
	\multicolumn{1}{c|}{12} &
	\multicolumn{1}{c|}{4} &
	\multicolumn{1}{c|}{3} &
	\multicolumn{1}{c|}{3} &
	\multicolumn{1}{c|}{3} &
	\multicolumn{1}{c|}{12} &
	\multicolumn{1}{c|}{12} &
	\multicolumn{1}{c|}{12} &
	\multicolumn{1}{c|}{0} &
	\multicolumn{1}{c|}{16} &
	\multicolumn{1}{c|}{9} &
	\multicolumn{1}{c|}{9} &
	\multicolumn{1}{c|}{9} &
	\multicolumn{1}{c|}{6} &
	\multicolumn{1}{c|}{6} &
	\multicolumn{1}{c|}{6} &
	\multicolumn{1}{c|}{0} &
	\multicolumn{1}{c|}{16} &
	\multicolumn{1}{c|}{11} &
	\multicolumn{1}{c|}{10} &
	\multicolumn{1}{c|}{10} &
	\multicolumn{1}{c|}{4} &
	\multicolumn{1}{c|}{5} &
	5 \\ \hline
	\multicolumn{1}{|c|}{\textbf{Total (163/83)}} &
	\multicolumn{1}{c|}{53} &
	\multicolumn{1}{c|}{31} &
	\multicolumn{1}{c|}{11} &
	\multicolumn{1}{c|}{8} &
	\multicolumn{1}{c|}{8} &
	\multicolumn{1}{c|}{72} &
	\multicolumn{1}{c|}{74} &
	\multicolumn{1}{c|}{74} &
	\multicolumn{1}{c|}{2} &
	\multicolumn{1}{c|}{82} &
	\multicolumn{1}{c|}{21} &
	\multicolumn{1}{c|}{21} &
	\multicolumn{1}{c|}{21} &
	\multicolumn{1}{c|}{61} &
	\multicolumn{1}{c|}{61} &
	\multicolumn{1}{c|}{61} &
	\multicolumn{1}{c|}{0} &
	\multicolumn{1}{c|}{84} &
	\multicolumn{1}{c|}{21} &
	\multicolumn{1}{c|}{19} &
	\multicolumn{1}{c|}{19} &
	\multicolumn{1}{c|}{61} &
	\multicolumn{1}{c|}{63} &
	63 \\ \hline
	\multicolumn{1}{|c|}{\textbf{Precision* (\#1, \#5, \#20)}} &
	\multicolumn{8}{c|}{0.5760, 0.5827, 0.5827} &
	\multicolumn{8}{c|}{0.9683, 0.9683, 0.9683} &
	\multicolumn{8}{c|}{1, 1, 1} \\ \hline
	\multicolumn{1}{|c|}{\textbf{Accuracy* (\#1, \#5, \#20)}} &
	\multicolumn{8}{c|}{0.6168, 0.6325, 0.6325} &
	\multicolumn{8}{c|}{0.8614, 0.8614, 0.8614} &
	\multicolumn{8}{c|}{0.8735, 0.8855, 0.8855} \\ \hline
	\multicolumn{1}{|c|}{\textbf{F1-score* (\#1, \#5, \#20)}} &
	\multicolumn{8}{c|}{\textbf{0.6923, 0.7081, 0.7081}} &
	\multicolumn{8}{c|}{\textbf{0.8414, 0.8414, 0.8414}} &
	\multicolumn{8}{c|}{\textbf{0.8531, 0.8690, 0.8690}} \\ \hline
	\multicolumn{1}{|c|}{\textbf{VEC (15/ 9)}} &
	\multicolumn{1}{c|}{2} &
	\multicolumn{1}{c|}{4} &
	\multicolumn{1}{c|}{3} &
	\multicolumn{1}{c|}{3} &
	\multicolumn{1}{c|}{3} &
	\multicolumn{1}{c|}{6} &
	\multicolumn{1}{c|}{6} &
	\multicolumn{1}{c|}{6} &
	\multicolumn{1}{c|}{0} &
	\multicolumn{1}{c|}{6} &
	\multicolumn{1}{c|}{3} &
	\multicolumn{1}{c|}{3} &
	\multicolumn{1}{c|}{3} &
	\multicolumn{1}{c|}{6} &
	\multicolumn{1}{c|}{6} &
	\multicolumn{1}{c|}{6} &
	\multicolumn{1}{c|}{0} &
	\multicolumn{1}{c|}{6} &
	\multicolumn{1}{c|}{9} &
	\multicolumn{1}{c|}{9} &
	\multicolumn{1}{c|}{9} &
	\multicolumn{1}{c|}{0} &
	\multicolumn{1}{c|}{0} &
	0 \\ \hline
	\multicolumn{1}{|c|}{\textbf{TO (26/ 12)}} &
	\multicolumn{1}{c|}{9} &
	\multicolumn{1}{c|}{5} &
	\multicolumn{1}{c|}{1} &
	\multicolumn{1}{c|}{1} &
	\multicolumn{1}{c|}{1} &
	\multicolumn{1}{c|}{11} &
	\multicolumn{1}{c|}{11} &
	\multicolumn{1}{c|}{11} &
	\multicolumn{1}{c|}{0} &
	\multicolumn{1}{c|}{14} &
	\multicolumn{1}{c|}{3} &
	\multicolumn{1}{c|}{3} &
	\multicolumn{1}{c|}{3} &
	\multicolumn{1}{c|}{9} &
	\multicolumn{1}{c|}{9} &
	\multicolumn{1}{c|}{9} &
	\multicolumn{1}{c|}{0} &
	\multicolumn{1}{c|}{14} &
	\multicolumn{1}{c|}{4} &
	\multicolumn{1}{c|}{4} &
	\multicolumn{1}{c|}{4} &
	\multicolumn{1}{c|}{8} &
	\multicolumn{1}{c|}{8} &
	8 \\ \hline
	\multicolumn{1}{|c|}{\textbf{Total (204/104)}} &
	\multicolumn{1}{c|}{64} &
	\multicolumn{1}{c|}{40} &
	\multicolumn{1}{c|}{15} &
	\multicolumn{1}{c|}{12} &
	\multicolumn{1}{c|}{12} &
	\multicolumn{1}{c|}{89} &
	\multicolumn{1}{c|}{91} &
	\multicolumn{1}{c|}{91} &
	\multicolumn{1}{c|}{2} &
	\multicolumn{1}{c|}{102} &
	\multicolumn{1}{c|}{27} &
	\multicolumn{1}{c|}{27} &
	\multicolumn{1}{c|}{27} &
	\multicolumn{1}{c|}{76} &
	\multicolumn{1}{c|}{76} &
	\multicolumn{1}{c|}{76} &
	\multicolumn{1}{c|}{0} &
	\multicolumn{1}{c|}{104} &
	\multicolumn{1}{c|}{34} &
	\multicolumn{1}{c|}{32} &
	\multicolumn{1}{c|}{32} &
	\multicolumn{1}{c|}{69} &
	\multicolumn{1}{c|}{71} &
	71 \\ \hline
	\multicolumn{1}{|c|}{\textbf{Precision (\#1, \#5, \#20)}} &
	\multicolumn{8}{c|}{0.5806, 0.5886, 0.5886} &
	\multicolumn{8}{c|}{0.975, 0.975, 0.975} &
	\multicolumn{8}{c|}{1, 1, 1} \\ \hline
	\multicolumn{1}{|c|}{\textbf{Accuracy (\#1, \#5, \#20)}} &
	\multicolumn{8}{c|}{0.6202, 0.6346, 0.6346} &
	\multicolumn{8}{c|}{0.8654, 0.8654, 0.8654} &
	\multicolumn{8}{c|}{0.8365, 0.8462, 0.8462} \\ \hline
	\multicolumn{1}{|c|}{\textbf{F1-score (\#1, \#5, \#20)}} &
	\multicolumn{8}{c|}{\textbf{0.6950, 0.7099, 0.7099}} &
	\multicolumn{8}{c|}{\textbf{0.8478, 0.8478, 0.8478}} &
	\multicolumn{8}{c|}{\textbf{0.8023, 0.8182, 0.8182}} \\ \hline
\end{tabular}%
	}
	\caption{Statistical metric of \sys compared to \tsan~\cite{Tsan} and \archer~\cite{Archer} on DRB~\cite{DRB}. The values in brackets for each microbenchmark category represent the total number of tests with established \textit{true positives}. LD: loop dependency/anti-dependency, SYN: sync, EB: explicit barrier, OOB: out-of-bound, VEC: vectorization, TO: target offload, SH: shared heap, TS: thread specific. TP: true positives, FP: false positives, FN: false negatives, TN: true negative. \textit{Accuracy} represents the closeness of the tool's classification metric with the ground truth values. \textit{Precision} represents the closeness of the tool's classification metric with its previously reported values. \textit{F1-score} represents the weighted average of the tool's \textit{accuracy} and \textit{precision}. \textit{Precision*, accuracy*, and f1-score*} showcase the statistical metrics without considering target offload and vectorization cases. TP and FN are categorized per execution count (\#), where the value shows the number of test runs performed to detect the reported TP and FN values. Lower variance in the values between test runs is preferred, signifying lower impact due to variations in thread interleavings.}
	\label{tab:drb-statistics}
\end{table*}

\begin{figure}[tbph]
	\centering
	\includegraphics[width=0.70\linewidth]{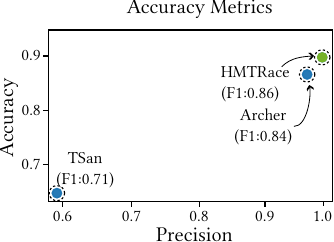}
	\captionof{figure}{Precision and accuracy of the dynamic race detection tools. The F1-score, representing the weighted average of precision and accuracy, is listed within parentheses.}
	\label{fig:drb-accuracy}
\end{figure}

\subsection{Precision}\label{sub:precision}

To evaluate the accuracy of race detection in \sys, we leverage \textit{DataRaceBench v1.4.1} (DRB) suite~\cite{DRB}, which provides an exhaustive list of curated microbenchmarks with established \textit{race ground-truth}: a subset of the microbenchmarks induce data races while providing a mechanism to statistically evaluate different race detection frameworks.

Table~\ref{tab:drb-statistics} shows a category-specific breakdown of the framework's precision metric, with figure~\ref{fig:drb-accuracy} showcasing the combined precision, accuracy, and f1-score of \sys against \tsan and \archer. The table showcases two separate categories of statistical metrics. Without considering $VEC$ and $TO$ categories of DRB, \sys has a  combined f1-score of \textit{0.86}, which is better than those of \tsan (\textit{0.71}) and \archer (\textit{0.84}). With these categories considered, \sys still exhibits a combined f1-score comparable to \archer. Furthermore, \sys does not report any false positives. A point to note is that while $TO$-based test cases support kernel offload to accelerators, the experiment uses existing CPU cores to execute the kernel.

\begin{figure*}[tbph]
	\centering
	\includegraphics[width=0.90\linewidth]{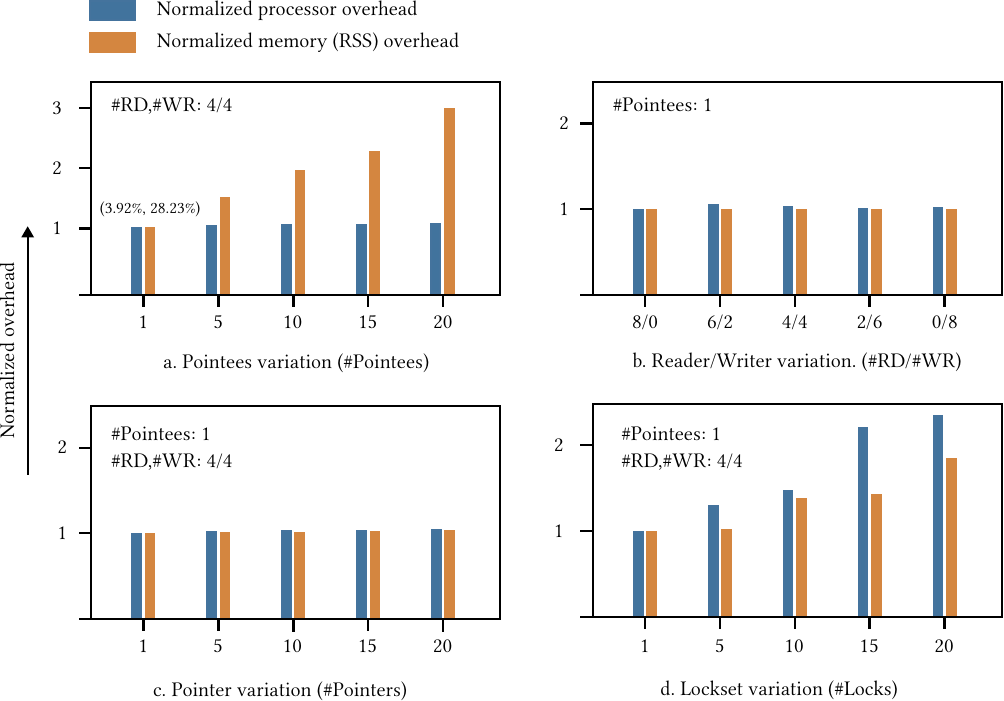}
	\captionof{figure}{Effect of varying number of (a) shared allocations, (b) reader/writer threads, (c) pointers, and (d) lock acquisitions. The experiment uses \textit{DRB075}~\cite{DRB} microbenchmark as a baseline. The value of the y-axis represents overheads (as a multiplier) normalized to the base case of (\#Pointee: 1,  \#RD/\#WR: 4/4) with a base value of \textit{3.92\%} processor overhead and \textit{28.23\%} memory (RSS) overhead when compared to non-instrumented binary.}
	\label{fig:causation}
\end{figure*}

\subsection{Execution time and memory overhead}\label{sub:execution-overhead}

Table~\ref{tab:execution-overhead} compares the execution time and peak memory (RSS) overhead of instrumentation employed in \sys against \tsan and \archer. The resource overhead for each benchmark is computed by considering vanilla non-instrumented binaries, compiled using clang (\textit{llvmorg-16.0.6})~\cite{LLVM:CGO04} as a baseline with \texttt{-O2} optimization set applied to both the non-instrumented and instrumented binaries. The overhead for \tsan and \archer are evaluated with the environment variable \textit{TSAN\_OPTIONS="report\_bugs=0"}, which sets the recommended configuration for benchmarking these frameworks~\cite{TsanOptions}: obtaining the overhead of instrumentation without the overhead of bug reporting.

\begin{table*}[tbph]
	\centering
	\resizebox{\textwidth}{!}{%
		\begin{tabular}{|c|c|ccc|ccc|}
			\hline
			\multirow{2}{*}{\textbf{Benchmark}} &
			\multirow{2}{*}{\textbf{\begin{tabular}[c]{@{}c@{}}Commit/\\ Version\end{tabular}}} &
			\multicolumn{3}{c|}{\textbf{Execution time overhead (\%)}} &
			\multicolumn{3}{c|}{\textbf{Peak memory (RSS) overhead (\%)}} \\ \cline{3-8} 
			&
			&
			\multicolumn{1}{c|}{\textbf{\begin{tabular}[c]{@{}c@{}}\tsan \\ (llvmorg-16.0.6)\end{tabular}}} &
			\multicolumn{1}{c|}{\textbf{\begin{tabular}[c]{@{}c@{}}\archer \\ (llvmorg-16.0.6)\end{tabular}}} &
			\textbf{\sys} &
			\multicolumn{1}{c|}{\textbf{\begin{tabular}[c]{@{}c@{}}\tsan \\ (llvmorg-16.0.6)\end{tabular}}} &
			\multicolumn{1}{c|}{\textbf{\begin{tabular}[c]{@{}c@{}}\archer \\ (llvmorg-16.0.6)\end{tabular}}} &
			\textbf{\sys} \\ \hline
			\textbf{DRB (avg)} &
			1.4.1 &
			\multicolumn{1}{c|}{438.29} &
			\multicolumn{1}{c|}{422.86} &
			6.28 &
			\multicolumn{1}{c|}{632.48} &
			\multicolumn{1}{c|}{560.13} &
			52.78 \\ \hline
			\textbf{Redis} &
			2962227 &
			\multicolumn{1}{c|}{512.35} &
			\multicolumn{1}{c|}{521.49} &
			4.81 &
			\multicolumn{1}{c|}{1121.59} &
			\multicolumn{1}{c|}{1068.2} &
			68.21 \\ \hline
			\textbf{Nginx} &
			f5a7f10 &
			\multicolumn{1}{c|}{384.71} &
			\multicolumn{1}{c|}{376.58} &
			3.74 &
			\multicolumn{1}{c|}{782.83} &
			\multicolumn{1}{c|}{761.99} &
			64.39 \\ \hline
			\textbf{Sqlite} &
			89efa89 &
			\multicolumn{1}{c|}{292.86} &
			\multicolumn{1}{c|}{284.32} &
			3.97 &
			\multicolumn{1}{c|}{794.26} &
			\multicolumn{1}{c|}{783.59} &
			62.9 \\ \hline
			\textbf{Memcached} &
			e0e415b &
			\multicolumn{1}{c|}{311.47} &
			\multicolumn{1}{c|}{301.74} &
			3.65 &
			\multicolumn{1}{c|}{752.88} &
			\multicolumn{1}{c|}{733.94} &
			69.63 \\ \hline
			\textbf{Pigz} &
			cb8a432 &
			\multicolumn{1}{c|}{368.92} &
			\multicolumn{1}{c|}{352.93} &
			3.92 &
			\multicolumn{1}{c|}{453.36} &
			\multicolumn{1}{c|}{481.16} &
			50.15 \\ \hline
			\textbf{Aget} &
			8fdc626 &
			\multicolumn{1}{c|}{480.46} &
			\multicolumn{1}{c|}{476.5} &
			4.07 &
			\multicolumn{1}{c|}{1017.11} &
			\multicolumn{1}{c|}{792.12} &
			61.06 \\ \hline
			\textbf{Apache Httpd} &
			cca42ea &
			\multicolumn{1}{c|}{391.26} &
			\multicolumn{1}{c|}{394.87} &
			3.62 &
			\multicolumn{1}{c|}{871.94} &
			\multicolumn{1}{c|}{721.58} &
			58.62 \\ \hline
			\textbf{Parsec.blackscholes} &
			\multirow{8}{*}{540b490} &
			\multicolumn{1}{c|}{231.01} &
			\multicolumn{1}{c|}{227.43} &
			3.44 &
			\multicolumn{1}{c|}{376.29} &
			\multicolumn{1}{c|}{292.84} &
			44.61 \\ \cline{1-1} \cline{3-8} 
			\textbf{Splash2x.fmm} &
			&
			\multicolumn{1}{c|}{286.39} &
			\multicolumn{1}{c|}{278.29} &
			3.58 &
			\multicolumn{1}{c|}{392.14} &
			\multicolumn{1}{c|}{298.57} &
			48.97 \\ \cline{1-1} \cline{3-8} 
			\textbf{Splash2x.ocean\_cp} &
			&
			\multicolumn{1}{c|}{329.22} &
			\multicolumn{1}{c|}{292.64} &
			3.85 &
			\multicolumn{1}{c|}{381.69} &
			\multicolumn{1}{c|}{284.38} &
			49.31 \\ \cline{1-1} \cline{3-8} 
			\textbf{Splash2x.radiosity} &
			&
			\multicolumn{1}{c|}{315.7} &
			\multicolumn{1}{c|}{336.97} &
			3.79 &
			\multicolumn{1}{c|}{414.38} &
			\multicolumn{1}{c|}{301.95} &
			54.68 \\ \cline{1-1} \cline{3-8} 
			\textbf{Splash2x.raytrace} &
			&
			\multicolumn{1}{c|}{388.96} &
			\multicolumn{1}{c|}{350.17} &
			3.91 &
			\multicolumn{1}{c|}{426.63} &
			\multicolumn{1}{c|}{412.6} &
			55.19 \\ \cline{1-1} \cline{3-8} 
			\textbf{Splash2x.volrend} &
			&
			\multicolumn{1}{c|}{364.95} &
			\multicolumn{1}{c|}{362.78} &
			3.98 &
			\multicolumn{1}{c|}{482.17} &
			\multicolumn{1}{c|}{391.25} &
			42.69 \\ \cline{1-1} \cline{3-8} 
			\textbf{Splash2x.water\_nsquared} &
			&
			\multicolumn{1}{c|}{437.62} &
			\multicolumn{1}{c|}{428.77} &
			4.02 &
			\multicolumn{1}{c|}{528.37} &
			\multicolumn{1}{c|}{421.45} &
			46.88 \\ \cline{1-1} \cline{3-8} 
			\textbf{Splash2x.water\_spatial} &
			&
			\multicolumn{1}{c|}{421.92} &
			\multicolumn{1}{c|}{431.8} &
			4.11 &
			\multicolumn{1}{c|}{564.19} &
			\multicolumn{1}{c|}{416.85} &
			48.35 \\ \hline
			\textbf{GEOMEAN} &
			\multicolumn{1}{l|}{} &
			\multicolumn{1}{c|}{\textbf{364.82}} &
			\multicolumn{1}{c|}{\textbf{356.94}} &
			\textbf{4.01} &
			\multicolumn{1}{c|}{\textbf{585.56}} &
			\multicolumn{1}{c|}{\textbf{500.19}} &
			\textbf{54.31} \\ \hline
		\end{tabular}%
	}
	\caption{Execution and peak memory (RSS) overhead of the instrumentations carried out by the race detection frameworks. Commit lists the benchmark versions used for evaluation. The values report percentage overhead when compared to vanilla non-instrumented binaries. The values shown are average overhead over 100 iterations.}
	\label{tab:execution-overhead}
\end{table*}

As seen in table~\ref{tab:execution-overhead}, the results showcase a general trend, where benchmarks instrumented with \sys race detection framework incur significantly lower execution time and memory overhead than \tsan and \archer. \tsan and \archer incur an average of \textit{364.82\%} and \textit{356.94\%} execution time overhead with an average of \textit{585.56\%} and \textit{500.19\%} additional physical memory usage (RSS) respectively. \sys, on the other hand, incurs an average overhead of \textit{4.01\%} for execution time and \textit{54.31\%} for physical memory usage. Between benchmarks, \sys exhibits lower execution time overhead on \textit{Sqlite} and \textit{Memcached} due to larger $READ/WRITE$ segments, which perform both temporal and spatial memory updates of the in-memory database. These benchmarks showcase lower transitions between segments, which translates to reduced instrumentation overhead of the tagging infrastructure for race detection.

\subsection{Causation analysis}\label{sub:causation}

Figure~\ref{fig:causation} provides insight into the overhead of the \sys framework with variance in the number of shared allocations, $RD$/$WR$ thread counts, number of pointers, and acquired locksets. We use \textit{DRB075}~\cite{DRB} as the baseline microbenchmark and introduce the necessary variations in its dataset to obtain causal insights. \sys applies similar instrumentation to both $RD$ and $WR$ segments, incurring similar processor and memory overhead penalties with varying numbers of $RD/WR$ counts. \sys does not observe considerable variance in overhead when increasing the number of pointers, as higher overhead instrumentations such as lockset updates and checks are performed only during segment start, with subsequent updates to pointer address incurring a minimal tag update overhead (same order as a volatile memory write). \sys only incurs memory overhead when increasing the number of pointees, as it has to align and pad these allocations while tracing lockset information for each pointee. \sys incurs both processor and memory overhead with an increase in lockset acquisition by $RD/WR$ threads due to the synchronization overhead of maintaining internal data structure for updating the locksets.

\subsection{New data races in large-scale applications}\label{sub:new-race}

Using \sys, we were able to detect four new data races in \textit{Redis (commit:2962227)}~\cite{Redis} and \textit{jemalloc (v5.3.0)}~\cite{jemalloc}, through fuzzing (using \textit{AFL++~\cite{Afl}}) and runtest module execution (packaged as part of Redis). At the time of this evaluation, none of these races were documented or listed as issues in its source repository.

Of the four newly identified data races, three races pertain to reader-writer conflicts, observed between the main redis-server thread and background (\textit{bg\_call\_worker}) threads spawned by the test module when simulating events such as \textit{blocked and unresponsive clients}. While these three races do not indicate an issue in the Redis server, they expose issues in the testing framework, part of the redis repository.

The last detected data race is exposed when redis is compiled with \textit{jemalloc} enabled. This package is made available as part of the Redis repository and implements better fragmented-memory management. The specific race is observed during the server \textit{fast-path} initialization when a \textit{cache\_bin} counter value is concurrently accessed without synchronization by the main redis-server thread and a background thread which is in the processor of returning a recently freed memory allocation to the cachec\_bin. A \textit{jemalloc} developer confirmed the validity of this race. They also commented that the counter value serves only to gather statistics around the cache\_bin usage, and a race in the counter would not affect the functionality of the redis-server. We confirm the reproducibility of these races with \tsan when fuzzing using the seed that exposes the racy interleaving.
\section{Limitations and future work}\label{sec:limitations}

This section addresses certain limitations of \sys and proposes future extensions to the framework.

\paragraph{\textbf{\mte intrinsic restrictions}}
\mte~\cite{MTE, MTEWhitepaper} is an optional processor intrinsic, part of \textit{ARMv8.5-A} specification. The 16B alignment and granule size requirements on the taggable allocations by \mte result in increased memory usage and fragmentation. A potential solution to work around this issue would be to extend the memory runtime allocators to manage tagged memory and allocate tagged-but-unused bytes to pointers that do not perform tagging. This would require propagating the tag values back to such pointers. We reserve this design change as future work.

\paragraph{\textbf{Limited \mte tag availability}}
\mte allows tagged pointers and their pointed-to allocations to one of fourteen tag values, with tag values of \textit{0b0000} and \textit{0b1111} reserved to represent non-tagged allocation and unfiltered tag access. While this limitation in the number of available tags does not affect the accuracy or precision of the \sys framework in detecting race, it does result in harder triages due to conflicts in the tags representing the racing thread IDs. On assertion of race due to \mte tag mismatch exception, the triage workflow would have to check for race between the faulting thread and all other threads with the same modulo-thread ID value.

\paragraph{\textbf{Vectorization and accelerator offload}}
Vectorization or SIMD instructions provide instruction-level parallelism, where a single vectorized instruction can operate on multiple contiguous data segments to perform parallel operations. The memory tag approach employed by \sys currently cannot detect races incurred through such vectorization instructions, as the approach would require additional insights into the targeted allocations before executing such instructions. 
We plan to assess the feasibility of vectorization support as part of future extensions to \sys. Furthermore, as memory tagging is supported only in anonymous pages, it cannot be leveraged to trace the data race between processor threads and accelerator kernels.

\paragraph{\textbf{Extensions to other parallel programming models}}
\sys currently does not support detecting data races in applications using object-oriented constructs. Analysis and instrumentation passes are not applied to object constructor, destructor, and virtual method functions. We reserve this design change as future work.
\section{Related work}\label{sec:related}

In this section, we review some prior works in data race detection that are not quantitatively evaluated in section~\ref{sec:eval}.

\paragraph{\textbf{Static race detection}} Race detection frameworks such as \textit{O2}~\cite{O2}, \textit{OpenRace}~\cite{OpenRace}, \textit{OMPRacer}~\cite{OMPRacer}, \textit{LLOV}~\cite{LLOV}, and \textit{RacerMod}~\cite{Racemob} have recently seen an increased emphasis in the scientific community due to the \textit{static} approach taken by these frameworks. Static analysis can be performed offline, on a more compute-intensive node, without executing the program being analyzed. The capability of analyzing programs without execution makes such approaches lucrative for analyzing programs with execution interleaving that are not regularly exercised during program execution. While these static approaches typically incur far fewer false negatives than most dynamic race detection frameworks, they suffer from state explosion problems~\cite{StaticDataRace}, usually encountered in large-scale programs, since precise race detection is considered to be a NP-hard problem~\cite{DataRace}. They also require accurate modeling of the execution model, which limits their portability and practicality, as is evident from the current industry adoption rate of static frameworks.

\paragraph{\textbf{Dynamic race detection}}\label{sub:related-dyn} Dynamic race detection frameworks provide a practical approach to race detection, albeit with a slightly higher false negative rate than static race detection frameworks. Frameworks such as \textit{Fastrack}~\cite{Fastrack}, \textit{Eraser}~\cite{Eraser}, \textit{Literace}~\cite{Literace}, \textit{Multirace}~\cite{Multirace}, and  \textit{Pacer}~\cite{Pacer} sample memory accesses and rely on \textit{happens-before} and \textit{lockset} analysis to determine potential data races. \tsan~\cite{Tsan} and \archer~\cite{Archer} derive inspiration from such frameworks and have significantly reduced the memory overhead of tracing memory accesses through the use of lightweight vector clocks (VCs). While the aforementioned dynamic race detection tools employ compile-time instrumentation to detect data races during the execution of the program, another set of race detection tools~\cite{Racez, Prorace} take a hybrid approach to trace memory during runtime and infer data race offline, after the program's execution. Happens-before~\cite{HBR} and lockset~\cite{LocksetAnalysis}  based approaches incur significant execution and memory overhead owing to the heavy instrumentation necessary to trace the allocation access. Sampling allocations instead of tracing every access may reduce the overhead, but it increases false negatives, making this approach less complete.
A framework that trades completeness for performance is \textit{IFRit}~\cite{Ifrit}, which enforces a dynamic sampling-based policy to coalesce memory access information to the same variable. \textit{TSVD}~\cite{TSVD} takes a different approach to race detection by leveraging continuous integration/deployment (CI/CD) frameworks to prune racy events through delay injection. By exploring the causality of delay injection, the framework is capable of inferring happens-before between operations and thereby pruning racy sets between sets of CI/CD runs.
	
\paragraph{\textbf{Hardware-assisted race detection}} An approach to reducing the execution and memory overhead inherent in dynamic race detection frameworks is to leverage hardware features available in the architecture for race detection. \textit{Datacollider}~\cite{Datacollider} was the earliest framework that leveraged processor debug registers and delay injection to infer data race. This approach serves as an inspiration to frameworks such as \textit{KCSAN}~\cite{KCSAN} and \textit{DRDDR}~\cite{Drddr}. However, this approach does not scale for programs without continuous verification cycles, constituting most large-scale userspace programs. 

\textit{KARD}~\cite{Kard} and \textit{PUSh}~\cite{PUSH} leverage \textit{Intel MPK} memory protection keys which enforce page granular memory access protection to trace concurrent access. While such approaches incur lower overhead, they have to enforce an application-specific policy to determine allocations that need to be isolated. \textit{PUSh} requires manual effort from the programmer to annotate allocations, which must be protected from concurrent access. A drawback of \textit{KARD} is that the framework can only identify inconsistent lock usage (ILU) and relies on the assumption that a traceable shared allocation is correctly accessed in at least one critical section through correct synchronization primitives.

Frameworks such as \textit{TxRace}~\cite{Txrace}, \textit{Aikido}~\cite{AIKIDO}, and \textit{PUSh}~\cite{PUSH} either use niche hardware modules or use custom hardware and hypervisor to trace memory accesses for a data race. \textit{Bugaboo}~\cite{Bugaboo} and frameworks based on \textit{Intel PT} are another set of race detection tools that leverage cache and processor tracing features to infer race. While tracing memory through dedicated hardware modules and cache is a suitable approach, capable of achieving higher accuracy and precision, they are less scalable and require more profound domain expertise to infer race from offline analysis.

For the evaluation of \sys, we chose \tsan and \archer, as these state-of-the-art dynamic race detection
frameworks support \textit{ARM} architecture. We could not evaluate against other hardware-based dynamic race detection frameworks, such as \textit{KARD}, as their artifacts are private~\cite{kardPrivate}. \textit{Datacollider}, while publicly available, is a framework for detecting data races in the Linux kernel and does not have a public port that supports userspace processes.
\section{Conclusion}
This paper presents \sys, a novel Armv8.5-A memory tag extension based dynamic data race detection framework. \sys can detect data races in multi-threaded applications based on \textit{OpenMP-} and \textit{Pthread-} shared memory programming models. Our experimental evaluations show that mainstream dynamic race detection tools such as \tsan and \archer, while accurate at detecting and reporting data races, incur substantial execution and memory overhead while reporting false positives. \sys, meanwhile, showcases comparable accuracy without reporting any false positives, all while incurring significantly lower execution overhead than both \tsan and \archer.


\bibliographystyle{plain}
\bibliography{references}

\end{document}